\documentclass[12pt]{iopart}
\usepackage[dvips]{graphicx}
\usepackage[latin1]{inputenc}
\usepackage{bbm}
\usepackage{psfrag}
\usepackage{iopams}

\newcounter{uebertrag}

\newcommand{\beq}{\begin{equation}} 
\newcommand{\eeq}{\end{equation}}
\newcommand{\bea}{\begin{eqnarray}} 
\newcommand{\eea}{\end{eqnarray}}

\newcommand{\RR}{{\mathbbm R}} 
\newcommand{\NN}{{\mathbbm N}}

\newcommand{\Hess}{{\mbox{Hess}}} 

\renewcommand{\labelenumi}{(\roman{enumi})}
 
\begin{document} 
 
\title{Kinetic energy and microcanonical nonanalyticities in finite and infinite
systems} 

\author{Lapo Casetti$^{1,2}$, Michael Kastner$^{3}$ and Rachele Nerattini$^1$}
\address{$^1$ Dipartimento di Fisica, Universit\`a di 
Firenze, via G.~Sansone 1, 50019 Sesto Fiorentino (FI), Italy}
\address{$^2$ Centro per lo Studio delle Dinamiche Complesse (CSDC), Universit\`a di 
Firenze, and Istituto Nazionale di Fisica Nucleare (INFN), Sezione di Firenze, 
via G.~Sansone 1, 50019 Sesto Fiorentino (FI), Italy}
\address{$^3$ National Institute for Theoretical Physics (NITheP), Stellenbosch 7600, South Africa}
\ead{\mailto{lapo.casetti@unifi.it},\mailto{kastner@sun.ac.za}}
 
\begin{abstract} 
In contrast to the canonical case, microcanonical thermodynamic func\-tions can show nonanalyticities also for finite systems. In this paper we contribute to the understanding of these nonanalyticities by working out the relation between nonanalyticities of the microcanonical entropy and its configurational counterpart. If the configurational microcanonical entropy $\omega_N^c(v)$ has a nonanalyticity at $v=v_c$, then the microcanonical entropy $\omega_N(\varepsilon)$ has a nonanalyticity at the same value $\varepsilon=v_c$ of its argument for any finite value of the number of degrees of freedom $N$. The presence of the kinetic energy weakens the nonanalyticities such that, if the configurational entropy is $p$ times differentiable, the entropy is $p+\lfloor N/2 \rfloor$-times differentiable. In the thermodynamic limit, however, the behaviour is very different: The nonanalyticities do not longer occur at the same values of the arguments, but the nonanalyticity of the microcanonical entropy is shifted to a larger energy. These results give a general explanation of the peculiar behaviour previously observed for the mean-field spherical model. With the hypercubic model we provide a further example illustrating our results.
\end{abstract} 
 
\pacs{05.20.Gg, 05.20.-y, 05.70.Fh} 
 
\section{Introduction}

In equilibrium statistical mechanics, nonanalyticities of thermodynamic functions are associated with thermodynamic phase transitions: more precisely, one commonly {\em defines} a phase transition point as the value of an external parameter (like temperature or pressure) where some thermodynamic function is nonanalytic.\footnote{Departing slightly from the standard definition, we use the notion of analyticity in the sense of a real function being infinitely-many times differentiable. This property typically, but not always coincides with the standard definition via the existence of a Taylor series.} Such an identification is satisfactory in the canonical ensemble: as originally suggested by Kramers \cite{Cohen:90}, nonanalyticities of thermodynamic functions calculated in the canonical ensemble may show up only in the thermodynamic limit $N\to\infty$, where $N$ is the number of degrees of freedom \cite{Griffiths}. Moreover, such nonanalyticities separate different phases, i.e., regions of the parameters where the collective properties of the system are different. When studying physical models, one usually finds that thermodynamic functions have only a small number of nonanalyticities. 

In the microcanonical ensemble, however, the situation is different. First, nonanalyticities of the microcanonical entropy may even be present at finite $N$. In principle, this fact should have been known for a long time, because even one-degree-of-freedom systems like a simple pendulum or a particle in a double-well potential do show nonanalyticities of the entropy. Still, it came as a surprise to many researchers in the field to see how frequently such nonanalyticities are encountered in many-particle systems \cite{KaSchne:06,DunHil:06,HilDun:06,CaKa:06}. The behaviour of such nonanalyticities as a function of the number $N$ of degrees of freedom is remarkable: their number may grow with $N$ even exponentially, and their ``strength'' generically decreases linearly with $N$. The latter means that the first $n$ derivatives of the entropy are continuous, where $n$ is of order $N$ \cite{KaSchneSchrei:07,KaSchneSchrei:08}. Since the usual thermodynamic quantities of interest are given by low-order derivatives of the entropy, these nonanalyticities of order $N$ can be observed only for very small $N$ from noisy data. In the thermodynamic limit most of these nonanalyticities disappear. Those (if any) that survive are typically associated to thermodynamic phase transitions and coincide with the canonical nonanalyticities if equivalence of statistical ensembles holds. Therefore the finite-$N$ nonanalyticities are not easily associated to any phenomenon that one would call a phase transition in the usual sense, and it seems inappropriate to define phase transitions in the microcanonical ensemble only based on the presence of nonanalyticities of the microcanonical entropy.

The general relation between nonanalyticities of the microcanonical entropy and thermodynamic phase transitions, i.e., the question which of the finite-system nonanalyticities survive in the thermodynamic limit, is a subject of active research (see e.g.\ \cite{Kastner:08,Pettini}). It remains an open problem, despite some recent advances where---under suitable conditions---the ``flatness'' of stationary points was shown to be relevant to whether its thermodynamic limit contribution is vanishing or not \cite{KaSchne:08,KaSchneSchrei:08}. In the present paper we will investigate the effect of a kinetic energy term on these nonanalyticities. The general properties of microcanonical nonanalyticities have been studied so far only when no kinetic term is present and the Hamiltonian is identified with the interaction potential energy. Standard kinetic energy terms in the Hamiltonian, i.e., quadratic forms in the momenta, are known to yield only trivial contributions to thermodynamics in the canonical ensemble. In the microcanonical ensemble a standard kinetic energy term may have a more noticeable effect: for instance, it may restore equivalence between canonical and microcanonical ensembles when only partial equivalence holds in the absence of a kinetic energy \cite{CaKa:07}. But adding a kinetic energy term has also a remarkable effect on nonanalyticities of the microcanonical entropy: For the exactly solvable mean-field spherical model it has been shown recently that, in the presence of a standard kinetic energy term, a nonanalyticity of the entropy which for any finite $N$ is located at a fixed value of the energy per degree of freedom, jumps discontinuously to a different value of the energy in the thermodynamic limit \cite{CaKa:06}.

In this paper we argue that such a behaviour of nonanalyticities in the ther\-mo\-dynamic limit is not a peculiarity of that model, but rather a general property of all the nonanalyticities that survive as $N \to \infty$. Moreover, we discuss the behaviour of nonanalyticities at finite $N$, showing that the kinetic energy weakens them: If a nonanalyticity of order $m$ is present in the configurational entropy, adding a kinetic energy term to the Hamiltonian will increase the order roughly to $m+N/2$, i.e., the first $m+N/2$ derivatives of the entropy will be continuous.

The paper is organized as follows. After giving some definitions and fixing some notation, in section \ref{sing} the behaviour of microcanonical nonanalyticities is discussed in general. More specifically, in section \ref{conf} we recall the results on nonanalyticities of the configurational entropy, while section \ref{kin} is devoted to the effect of the kinetic energy. In section \ref{hyper} we present a simple example to illustrate this behaviour. We will finish with some concluding remarks in section \ref{concl}.

\section{Microcanonical entropy and its nonanalyticities} 
\label{sing}

We consider classical Hamiltonian systems with $N$ degrees of freedom, with Hamiltonian function ${\cal H}:\Lambda_N\mapsto\RR$ of the form
\beq
\label{H}
{\mathcal H}(p,q)=\frac{1}{2}\sum_{i=1}^N p_i^2 + V(q)
\eeq
with some potential energy $V:\Gamma_N\mapsto\RR$. We denote by $\Gamma_N\subseteq\RR^N$ the $N$-dimensional configuration space and by $\Lambda_N$ its cotangent bundle, i.e., the phase space. We shall usually denote configurations as $q=(q_1,\ldots,q_N)\in\Gamma_N$ and phase space points as $(p,q)=(p_1,\ldots,p_N,q_1,\ldots,q_N)\in\Lambda_N$. 

The fundamental quantity of the microcanonical ensemble is the {\em microcanonical entropy}\/ as a function of the energy (per degree of freedom) $\varepsilon$,
\begin{equation}
\label{s}
s_N(\varepsilon)= \frac{1}{N}\ln \omega_N(\varepsilon),
\end{equation}
where
\beq
\label{omega}
\omega_N(\varepsilon) = \int_{\Lambda_N} \rmd  p\, \rmd  q\, \delta\left[{\cal H}(p,q)-N\varepsilon\right]
\eeq
is the {\em density of states} and $\delta$ denotes the Dirac distribution\footnote{We define all thermodynamic functions {\em per degree of freedom}, which accounts for the factor $1/N$ in the definitions. The Boltzmann constant $k_{\mathrm{B}}$ is set to unity.}. A related quantity is the {\em configurational microcanonical entropy}\/ as a function of the potential energy (per degree of freedom) $v$,
\begin{equation}\label{eq:s_conf}
s^{\mathrm{c}}_N(v)=\frac{1}{N}\ln \omega^c_N(v),
\end{equation}
where
\beq
\label{omegac}
\omega^c_N(v) = \int_{\Gamma_N} \rmd  q\, \delta\left[V(q)-Nv\right]
\eeq
is the {\em configurational density of states}. The configurational entropy equals the entropy when the Hamiltonian just consists of a configuration-dependent potential energy, ${\cal H} \equiv V$. This is often the case when studying discrete spin systems where a definition of conjugate momenta is difficult.

Alternatively, one can define the {\em integrated} density of states
\beq
\label{intomega}
\Omega_N(\varepsilon) = \int_{\Lambda_N} \rmd  p\, \rmd  q\, \Theta\left[N\varepsilon - {\cal H}(p,q)\right],
\eeq
where $\Theta$ is the Heaviside step function, and the corresponding entropy function is
\begin{equation}
\label{sigma}
\sigma_N(\varepsilon)= \frac{1}{N}\ln \Omega_N(\varepsilon).
\end{equation}
Again, the configurational counterparts can be defined, where
\beq
\label{intomegac}
\Omega^c_N(v) = \int_{\Gamma_N} \rmd  q\, \Theta\left[Nv - V(q)\right]
\eeq
and the entropy is given by
\begin{equation}
\sigma^c_N(v)= \frac{1}{N}\ln \Omega^c_N(v).
\end{equation}
The density of states \eref{omega} and the integrated density of states \eref{intomega} are related by 
\beq
\omega_N(\varepsilon) = \frac{\rmd \Omega_N}{\rmd \varepsilon}, \label{deriv}
\eeq
and an analogous relation holds for the configurational quantities $\Omega_N^c$ and $\omega_N^c$. Under suitable conditions on the Hamiltonian the difference between the entropies vanishes in the thermodynamic limit,
\beq
s_\infty \equiv \lim_{N\to\infty} s_N = \lim_{N\to\infty} \sigma_N,\qquad s^c_\infty \equiv \lim_{N\to\infty} s^c_N = \lim_{N\to\infty} \sigma^c_N
\eeq
(see section 3.3.14 of \cite{Ruelle} for details). Unless explicitly noted, the general properties of entropy functions considered in the following hold true for both definitions \eref{s} and \eref{sigma}.

As mentioned before, the microcanonical entropy is not necessarily an analytic function, neither for finite $N$ nor in the thermodynamic limit. We will point out in the following that nonanalyticities of the entropy originate from {\em stationary points} of the Hamiltonian \cite{KaSchneSchrei:08}. A stationary point of a function $f: M \subseteq \RR^n \mapsto \RR$ is a point $x_c \in M$ such that $\rmd f(x_c) = 0$, and the value $f(x_c)$ is called a {\em stationary value} of $f$. When stationary points are {\em non-degenerate}, i.e., the Hessian matrix $\Hess_f$ of $f$ is nonsingular at all stationary points $x_c$, the function $f$ is called a {\em Morse function}. In this case all the stationary points are isolated. The {\em index} $j$ of the stationary point $x_c$ is the number of negative eigenvalues of the Hessian at $x_c$. Minima and maxima are stationary points corresponding to $j = 0$ and $j = n$, respectively.

In the following we will assume the potential energy $V(q)$ to be a Morse function unless explicitly stated otherwise. Conceptually, this is an insignificant restriction, since Morse functions on some manifold $M$ form an open dense subset in the space of smooth functions on $M$ \cite{Demazure}, and are therefore generic. Hence, if the potential is not a Morse function, we can deform it into a Morse function by adding an arbitrarily small perturbation.\footnote{In case $V$ is not a Morse function due to a continuous symmetry, Morse-Bott theory \cite{Frankel} should allow to carry over essentially all the results from the theory of standard Morse functions.}

If the Hamiltonian ${\cal H}(p,q)$ is of standard form \eref{H}, its stationary points are of the form
\beq
\label{Hcritpoints}
(p_c,q_c) = (0,q_c), 
\eeq
where $q_c$ is a stationary point of the potential energy $V(q)$. Hence the stationary value of the Hamiltonian coincides with the stationary value of the potential energy, i.e., $\varepsilon_c = v_c$. Let us first recall some results on the relation between the nonanalyticities of the {\em configurational} entropy $s_N^c$ and the stationary points $q_c$ of $V$.

\subsection{Nonanalyticities of the configurational entropy}
\label{conf}

The Morse property of the potential $V$ ensures that its stationary points are isolated, so that we can safely restrict our attention to a single stationary point $q_c$. We denote by $v_c=V(q_c)/N$ the corresponding stationary value per degree of freedom. It has been shown in \cite{KaSchneSchrei:08} that in this situation the configurational density of states (\ref{omegac}) can be written, to the leading order, as
\beq
\label{omegacrit}
\omega^c_N(v) = P(v-v_c) + \frac{h_{N,j}(v-v_c)}{\sqrt{\left|\det \left[\Hess_V(q_c)
\right] \right|}} + o\left[(v-v_c)^{N/2 - \delta} \right]
\eeq
with some $\delta > 0$, where $P$ is a polynomial of degree smaller than $N/2$ in $v-v_c$, and
\beq
\label{hcrit}
h_{N,j}(x) = \cases{
(-1)^{j/2} x^{(N-2)/2} \Theta(x) & \mbox{for $j$ even},\\ 
(-1)^{(j+1)/2} x^{(N-2)/2} \pi^{-1} \ln |x| & \mbox{for $N$ even, $j$ odd},\\ 
(-1)^{(N - j)/2} (-x)^{(N-2)/2} \Theta(-x) & \mbox{for $N$, $j$ odd}.
}
\eeq
If there are further stationary points, the configurational density of states is given by the sum of the contributions of each stationary point. A proof (of an even stronger result including higher order terms) is given in \cite{KaSchneSchrei:08}. This result can be rephrased as follows:
\begin{enumerate}
\renewcommand{\labelenumi}{(\roman{enumi})}
\item Every stationary point $q_c$ of $V$ gives rise to a nonanalyticity of the configurational entropy $s_N^c(v)$ at the corresponding stationary value $v = v_c = V(q_c)/N$.
\item The order of this nonanalyticity is $\lfloor(N - 3)/2\rfloor$, i.e., $s_N^c(v)$ is precisely $\lfloor(N - 3)/2\rfloor$-times differentiable at $v = v_c$.\footnote{With $\lfloor x \rfloor$ we denote the largest integer smaller than $x$.}
\setcounter{uebertrag}{\value{enumi}}
\addtocounter{uebertrag}{-1}
\end{enumerate}
Since the integrated density of states $\Omega^c_N$ is obtained from $\omega_N^c$ by integration, the nonanalyticity of the entropy following from definition (\ref{sigma}) is slightly weaker. In this case, statement (ii) has to be replaced by:
\begin{enumerate}
\renewcommand{\labelenumi}{(\roman{enumi}$'$)}
\setcounter{enumi}{\value{uebertrag}}
\item The configurational entropy $\sigma_N^c(v)$ is precisely $\lfloor(N - 1)/2\rfloor$-times differentiable at $v = v_c$.
\end{enumerate}

\subsection{The role of kinetic energy}
\label{kin}

We have pointed out in equation \eref{Hcritpoints} that, for standard Hamiltonians of the form \eref{H}, if $q_c$ is a stationary point of the potential energy $V(q)$ then $(0,q_c)$ is a stationary point of ${\cal H}(p,q)$ and {\em vice versa}. Hence the kinetic energy is zero at stationary points and ${\cal H}(p_c,q_c)=V(q_c)$ for all stationary points $(p_c,q_c)$ of ${\cal H}$. As a consequence, for all finite $N$ the nonanalyticities of the configurational entropy---which we have traced back to stationary points in the previous section---show up at the very same stationary values as those of the entropy. As we shall see below, the presence of a kinetic energy term has a twofold effect on these nonanalyticities.
\begin{enumerate}
\renewcommand{\labelenumi}{\arabic{enumi}.}
\item At any finite $N$, the order of the nonanalyticity is increased by the presence of a kinetic energy term.
\item More surprisingly, in the thermodynamic limit those of the nonanalyticities which survive jump to a different value of the energy. 
\end{enumerate}
Both these results follow from the fact that for Hamiltonians of the class (\ref{H}) the density of states can be written as a convolution \cite{Ruelle}. Defining a {\em kinetic density of states} as
\beq
\label{omegak}
\omega^k_N(\gamma) = \int_{\RR^N} \rmd p \,
\delta\Biggl(\frac{1}{2}\sum_{i=1}^N p_i^2-N\gamma\Biggr),
\eeq
we can write the density of states as
\beq
\label{conv}
\omega_N(\varepsilon) = \int_0^{\infty} \rmd\gamma \, \omega^k_N(\gamma)\omega^c_N(\varepsilon - \gamma)
 = \int_{-\infty}^\varepsilon \rmd\gamma \, \omega^k_N(\varepsilon-\gamma)\omega^c_N(\gamma),
\eeq
where the configurational density of states $\omega^c_N$ is given by equation \eref{omegac}.

Since equations \eref{omegak} and \eref{conv} hold in the same form also for the integrated densities of states $\Omega$, all the following results will be valid also for integrated densities of states and entropies $\sigma$ defined as in equation \eref{sigma}. We will now discuss the finite-$N$ case and the thermodynamic limit separately in the following two subsections. 

\subsubsection{The finite-$N$ case.}
\label{finiten}

At any finite $N$, the effect of the kinetic energy term on the order of the nonanalyticities of the entropy can be computed explicitly from the convolution integral \eref{conv}. Such a calculation is reported in \ref{appendix}, and the only additional input used is that---in accordance with the results from section \ref{conf}---the nonanalyticities of the configurational density of states $\omega^c_N$ are of algebraic type. 

Alternatively, the result of that calculation can be anticipated via an intuitive argument: We know that $\omega_N^k$ is smooth, and we assume for simplicity that the configurational density of states is nonanalytic only at $v = v_c$ and analytic elsewhere. As long as $\varepsilon< v_c$, it is evident from the right-hand side of \eref{conv} that the nonanalyticity of $\omega^c_N$ at $v_c$ is never reached in the integration. Hence the convolution integral (\ref{conv}) is the integral over the product of two analytic functions and yields an analytic function. As soon as $\varepsilon > v_c$, the nonanalyticity of $\omega^c_N$ is inside the range of integration and induces a nonanalyticity in the convolution integral. As a simple example consider the convolution integral \eref{conv} with the choices $\omega_N^k(x) = x$ and $\omega_N^c(x) = \Theta(x-a)$ with $a>0$. Performing the integration yields
\beq
\label{simple}
\omega_N(\varepsilon) = \frac{1}{2}(\varepsilon - a)^2 \,\Theta(\varepsilon-a).
\eeq
As expected from the above reasoning, this function inherits the nonanalyticity at $\varepsilon = a$ from $\omega_N^c$. This is in agreement with our previous observation that the stationary values of the Hamiltonian coincide with those of the potential energy. In this example, $\omega_N^c$ is a 0-times differentiable function, whereas $\omega_N$ is 1-times differentiable. The calculation in \ref{appendix} shows that in general adding a kinetic energy increases the order of the nonanalyticity.

The effect of a standard kinetic energy term on the nonanalyticities of the entropy at finite $N$, as computed in \ref{appendix}, can be summarized as follows:
\begin{enumerate}
\renewcommand{\labelenumi}{(\roman{enumi})}
\item If the configurational density of states $\omega^c_N(v)$ is nonanalytic at $v =v_c$, then the density of states $\omega_N(\varepsilon)$ and the entropy $s_N(\varepsilon)$ are nonanalytic at $\varepsilon =v_c$;
\item 
The density of states $\omega_N(\varepsilon)$ and the entropy $s_N(\varepsilon)$ at $\varepsilon= v_c$ are differentiable $\lfloor N/2\rfloor$-times more often than the configurational density of states $\omega^c_N(v)$.
\setcounter{uebertrag}{\value{enumi}}
\addtocounter{uebertrag}{-1}
\end{enumerate}
As far as the entropy is concerned, statements (i) and (ii) above hold if $v_c$ is in the interior of the domain of the entropy. Both statements hold also for the integrated densities of states $\Omega_N(\varepsilon)$ and $\Omega^c_N(\varepsilon)$ and for the entropy $\sigma_N(\varepsilon)$. 

\subsubsection{Thermodynamic limit.}
\label{tdlimit}
We will assume in the following that the thermodynamic limit of the configurational entropy $s^c_N$ exists, i.e., that $\omega^c_N=\exp(Ns^c_N)$ increases exponentially with $N$ asymptotically for large $N$. From definition \eref{omegak} it follows that the kinetic density of states $\omega^k_N(\gamma)$ is related to the volume of an $(N-1)$-dimensional sphere with radius $\sqrt{2N\gamma}$,
\beq\label{eq:omegakin}
\omega^k_N(\gamma)=\frac{1}{\sqrt{2N\gamma}}\int_{\RR^N} \rmd p \,\delta\Biggl(\sqrt{\sum p_i^2}-\sqrt{2N\gamma}\Biggr)=a_N\gamma^{N/2-1}
\eeq
with
\beq
a_N=2\frac{\pi^{N/2}}{\Gamma(N/2)}(2N)^{N/2-1}.
\eeq
As a consequence, this quantity likewise grows exponentially in $N$ and the ther\-mo\-dy\-nam\-ic limit of the kinetic entropy $s_N^k=\ln(\omega^k_N)/N$ exist. Under these conditions, the thermodynamic limit is known to exist also for the entropy $s_N$ (see section 3.4.1 of \cite{Ruelle}) and is given, apart from irrelevant constants, by
\beq
\label{laplace}
s_\infty(\varepsilon) = \lim_{N\to\infty} \frac{1}{N} \ln\omega_N(\varepsilon)
= \lim_{N\to\infty} \frac{1}{N} \ln \max_{\gamma\geqslant0} 
\left[\omega^k_N(\gamma)\, \omega^c_N(\varepsilon - \gamma)\right]. 
\eeq
Denoting by $\tilde\gamma(\varepsilon)$ the value of $\gamma$ that realizes the extremum in the right-hand side of equation \eref{laplace}, this expression can be rewritten as
\beq
\label{sinfintermedio}
s_\infty(\varepsilon) = s^k_\infty[\tilde\gamma(\varepsilon)] +
s^c_\infty[\varepsilon - \tilde\gamma(\varepsilon)]. 
\eeq
From the above equation it is apparent that $\tilde\gamma(\varepsilon)$ must be equal to the microcanonical average of the kinetic energy per degree of freedom,
\beq\label{ekin}
\tilde\gamma(\varepsilon) = \frac{1}{N}\Biggl\langle 
\frac{1}{2}\sum_{i=1}^N p_i^2\Biggr\rangle
 = \varepsilon - \frac{1}{N}\left\langle V(q) \right\rangle
 = \varepsilon - \langle v \rangle(\varepsilon),
\eeq
which allows us to rewrite equation (\ref{sinfintermedio}) in the form 
\beq
\label{sfullinf}
s_\infty(\varepsilon) = s^k_\infty[\varepsilon - \langle v \rangle(\varepsilon)]+ s^c_\infty[\langle v \rangle(\varepsilon)]. 
\eeq
If $s^c_\infty(v)$ is nonanalytic at $v=v_c$, then $s_\infty(\varepsilon)$ will be nonanalytic at $\varepsilon=\varepsilon^\ast$, where $\varepsilon^\ast$ is defined implicitly by 
\beq\label{defestar}
\langle v \rangle(\varepsilon^\ast) = v_c.
\eeq
Apparently $\varepsilon^\ast$ differs from the value of $v_c$, unless the average kinetic energy \eref{ekin} vanishes at $\varepsilon^\ast$. Hence, despite their common origin from the nonanalyticity of $s^c$ at $v_c$, nonanalyticities of $s_\infty(\varepsilon)$ jump from their finite-$N$ value of $\varepsilon$ to a different value in the thermodynamic limit. 

These properties of the nonanalyticities of the microcanonical entropy at finite and infinite $N$ had previously been studied for a simple model system, the mean-field spherical model. The results reported in Refs.\ \cite{CaKa:06,CaKa:07}---surprising at the time---are all in agreement with and satisfactorily explained by the results reported in the present article. To further illustrate the predictions, we will discuss in the next section an even simpler, analytically solvable model which is of pedagogical value. 
 
\section{A simple example: the hypercubic model}
\label{hyper}

\begin{figure}[ht]
\center
\includegraphics[width=8cm,clip=true]{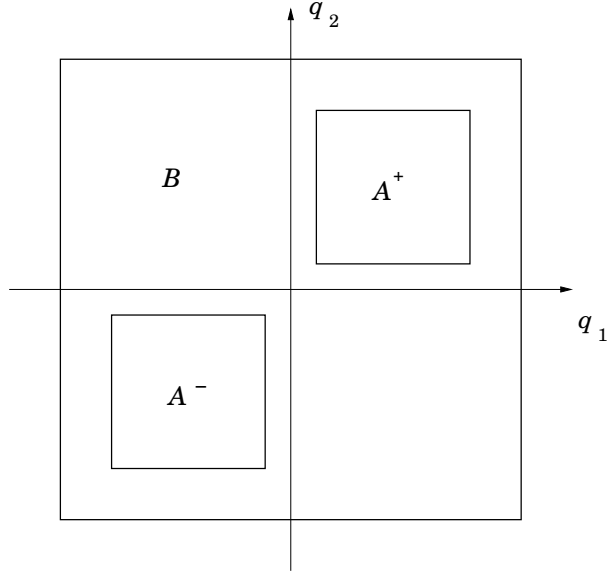}
\caption{Arrangement of hypercubes defining the potential energy \eref{pot1} of the hypercubic model for $N=2$.}
\label{ipercubi}
\end{figure}

The hypercubic model, introduced in \cite{BaCa:06}, can be seen as an $N$-dimensional gen\-er\-al\-i\-za\-tion of a particle in a one-dimensional potential made up of two square wells separated by a finite barrier. The Hamiltonian is of standard form \eref{H}, i.e., is given by a standard kinetic term plus a potential energy $V(q)$. To define the potential, we consider a hypercube $B$ of side length $b$, centered at the origin, as well as two disjoint hypercubes $A^{+}$, $A^{-}\subset B$ of side length $a\leqslant b/2$, symmetrically arranged with respect to the hyperplane $q_1+q_2+ \cdots + q_N = 0$ (see sketch in Fig.\ \ref{ipercubi}). The potential of the hypercubic model is then defined as
\begin{equation}\label{pot1}
V(q)= \cases{
0 & \mbox{for $q\in \{ A^+ \cup A^- \}$,} \\ 
Nv_c & \mbox{for $q\in B\setminus \{ A^+ \cup A^- \}$,}\\ 
\infty & \mbox{for $q\in \RR^N\setminus B$}. 
}
\end{equation}
The integrated density of states of this model can be computed analytically for arbitrary $N$, and the effect of a kinetic energy term will nicely illustrate the general results of the previous section.

\subsection{Finite-$N$ integrated microcanonical entropy}

The integrated density of states is calculated for the hypercubic model by inserting the potential \eref{pot1} into definition \eref{intomega},
\begin{eqnarray}
&\fl{\Omega_N(\varepsilon) = \int_{\RR^N} \rmd p \int_{A^+ \cup A^-} \rmd q\,\Theta\Biggl(N\varepsilon - \sum_{i=1}^N \frac{p_i^2}{2}\Biggr)} \nonumber \\ 
& + \int_{\RR^N} \rmd p \int_{B\setminus\{A^+ \cup A^- \}} \rmd q\,
\Theta\Biggl[N(\varepsilon - v_c)- \sum_{i=1}^N \frac{p_i^2}{2}\Biggr] \nonumber
\\ 
&\fl = 2a^N \int_{\RR^N} \rmd p \,
\Theta\Biggl(N\varepsilon - \sum_{i=1}^N \frac{p_i^2}{2}\Biggr) + \left(b^N - 2a^N\right) 
\int_{\RR^N} \rmd p \,
\Theta\Biggl[N(\varepsilon - v_c)- \sum_{i=1}^N \frac{p_i^2}{2}\Biggr].
\end{eqnarray}
The remaining integrations of the form
\beq
\label{volball}
\int_{\RR^N} \rmd p \,
\Theta\Biggl(\frac{\gamma^2}{2} - \sum_{i=1}^N \frac{p_i^2}{2}\Biggr)
= C_N\, \gamma^N\qquad \mbox{with}\;C_N = \frac{\pi^{N/2}}{\Gamma(N/2 + 1)}
\eeq
yield the volumes of $N$-dimensional balls of radii $\gamma=\sqrt{2N\varepsilon}$ and $\gamma=\sqrt{2N(\varepsilon-v_c)}$, respectively.
Inserting this formula into $\Omega_N$ and making use of \eref{sigma}, the integrated microcanonical entropy
\beq\label{sigmahyper}
\fl\sigma_N(\varepsilon) = 
\cases{
\left(\frac{1}{N}+\frac{1}{2}\right)\ln 2 +\ln a 
+ \frac{1}{N} \ln C_N 
+ \frac{1}{2}(\ln N + \ln \varepsilon) &  \mbox{for $0 < \varepsilon < v_c$},\\
{ \frac{1}{N} \ln C_N +
\frac{1}{N} \ln\left[2a^{N}(2N\varepsilon)^{N/2} +
(b^{N}-2a^{N})[2N(\varepsilon-v_{c})]^{N/2} \right]} & \mbox{for
$\varepsilon \geqslant v_c$},
}
\eeq
can be computed.
For the integrated configurational microcanonical entropy, a similar calculation yields
\begin{equation}\label{sigmahyperN}
\sigma^c_N(v)= \cases{
\ln a + \frac{1}{N} \ln 2 & \mbox{for $0< v < v_c$},\\
\ln b & \mbox{for $v \geqslant v_c$}.
}
\end{equation}
From these equations we see that $\sigma_N$ and $\sigma^c_N$ each have one nonanalyticity in the interior of their domains $[0,\infty)$, located at argument $v_c$ in both cases. The integrated configurational microcanonical entropy $\sigma^c_N$ is a piecewise constant function with a discontinuity\footnote{Since the potential \eref{pot1} is not a Morse function, it is no surprise that the nonanalyticity is not of the generic type given in equations \eref{omegacrit} and \eref{hcrit}.} at $v = v_c$,
whereas the 
functional form of the integrated microcanonical entropy $\sigma_N$ in the vicinity of $v_c$ is 
$\left | \varepsilon - v_c \right |^{N/2}$, 
in agreement with the general results of section \ref{finiten}.

\subsection{Thermodynamic limit}

\begin{figure}[ht]
\center
\psfrag{s}{$\sigma_N$}
\psfrag{e}{$\varepsilon$}
\includegraphics[width=12cm,clip=true]{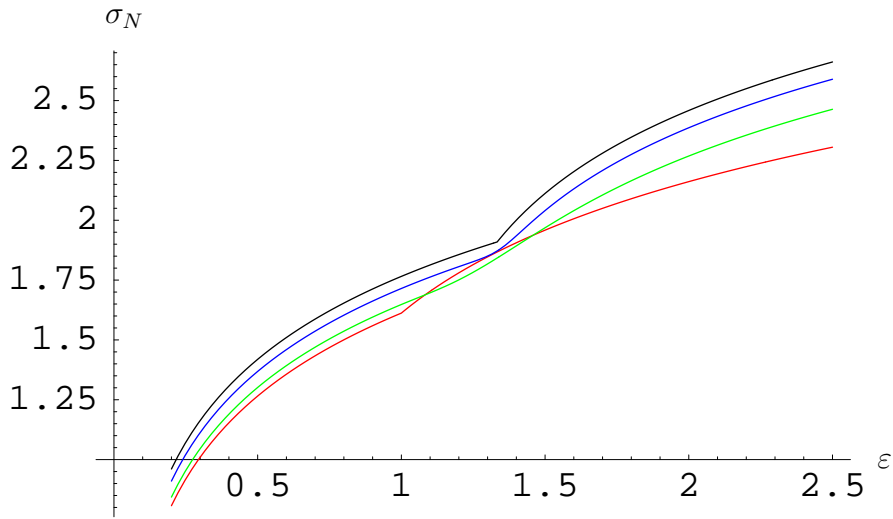}
\caption{Entropy $\sigma_N$ of the hypercubic model, plotted as a function of the energy $\varepsilon$ for $N= 2$ (red), $N = 8$ (green), $N = 32$ (blue) and $N = \infty$ (black). The numerical values of the parameters are $v_c = 1$, $a = 1$, $b = 2$, so that $\varepsilon^\ast = 4/3$.}
\label{entropiamicrocanonico}
\end{figure}

From the entropies \eref{sigmahyper} and \eref{sigmahyperN}, the corresponding thermodynamic limit values
\begin{equation}
\sigma_\infty(\varepsilon) = \lim_{N\to\infty}\sigma_N(\varepsilon)\qquad\mbox{and}\qquad\sigma_\infty^c(\varepsilon) = \lim_{N\to\infty}\sigma_N^c(\varepsilon)
\end{equation}
can be computed. Since
\beq
\lim_{N\to\infty}  \frac{1}{N} \ln C_N = \frac{1}{2}\left(1+ \ln \pi +\ln 2 - \ln N \right) ,
\eeq
for $0< \varepsilon < v_c$ we find
\begin{equation}\label{s1}
\sigma_{\infty}(\varepsilon)=\ln a + \frac{1}{2}\ln \varepsilon +
\frac{1}{2}\ln \pi + \ln 2 + \frac{1}{2},
\end{equation}
whereas for the case $\varepsilon\geqslant v_{c}$ we can write
\bea
\sigma_\infty(\varepsilon) & = & \frac{1}{2}(1 + \ln \pi +\ln 2-\ln N)\label{ss}\\
& + &
\lim_{N\to\infty}\frac{1}{N}\ln\left\{2a^{N}(2N\varepsilon)^{N/2}+(b^{N}-2a^{N})\left[(2N)^{N/2}(\varepsilon-v_{c})^{N/2}\right] \right\}. \nonumber
\eea
The argument of the logarithm in equation (\ref{ss}) is the sum of two terms. Both terms are exponentially large in $N$, so that in the limit $N\to\infty$ only the larger one survives. There is a value
\begin{equation}\label{estar}
\varepsilon^{\ast}=\frac{v_{c}}{1-\left( \frac{a}{b}\right)^{2}
},
\end{equation}
of $\varepsilon$ for which the two terms are equal, and apparently $\varepsilon^\ast > v_c$. For $\varepsilon < \varepsilon^{\ast}$, the first term in the argument of the logarithm survives, yielding the same functional form for $\sigma_\infty$ as in the case $0<\varepsilon < v_c$. For $\varepsilon \geqslant\varepsilon^{\ast}$, however, the second term wins, and we obtain as a final result
\beq\label{sigmainfhyper}
\fl\sigma_\infty(\varepsilon) = 
\cases{
{\frac{1}{2} + \ln 2 +\ln a + \frac{1}{2}\ln \pi
+ \frac{1}{2}\ln \varepsilon} & \mbox{for $0< \varepsilon < \varepsilon^\ast$},\\
{\frac{1}{2} + \ln 2 +\ln b + \frac{1}{2}\ln \pi
 + \frac{1}{2}\ln (\varepsilon - v_c) } & \mbox{for $\varepsilon \geqslant \varepsilon^\ast$}.\\
}
\eeq
This expression shows that the entropy $\sigma_\infty(\varepsilon)$ is analytic at $\varepsilon=v_c$ but nonanalytic at $\varepsilon = \varepsilon^\ast \neq v_c$. Therefore, although the entropy is nonanalytic at $\varepsilon = v_c$ for all finite $N$, the nonanalyticity jumps to a different energy value $\varepsilon^\ast$ in the thermodynamic limit. Furthermore, at this value the statistical average $\langle v\rangle$ of the potential energy per degree of freedom $v$ equals the value of the finite-$N$ nonanalyticity,
\beq
\langle v \rangle (\varepsilon^\ast) = v_c.
\eeq 
A plot of the entropy for various finite values of $N$ as well as in the thermodynamic limit is shown in Fig.\ \ref{entropiamicrocanonico}. Again, the results for the hypercubic model confirm the general reasoning of section \ref{tdlimit}.

From equation (\ref{sigmainfhyper}) the microcanonical temperature
\beq
T(\varepsilon) = \Bigl(\frac{\rmd s_N}{\rmd \varepsilon}\Bigr)^{-1} = 
\cases{
2\varepsilon & \mbox{for $\varepsilon < \varepsilon^\ast$},\\
2(\varepsilon -v_c) & \mbox{for $\varepsilon \geqslant \varepsilon^\ast$}.
}
\eeq
is easily computed. The jump in temperature at $\varepsilon^\ast$ is a signal of a microcanonical first-order phase transition.

\section{Concluding remarks}
\label{concl}

Previously, nonanalyticities of the microcanonical density of states and of the mi\-cro\-ca\-non\-i\-cal entropy usually had been investigated for {\em configurational} quantities. This choice was mostly motivated by the belief that the effect of a standard kinetic energy term, i.e., a quadratic form in the momenta, is trivial anyway. However, we have shown in the present article that the effect of a standard kinetic energy term on the nonanalyticities of the microcanonical entropy is quite remarkable, both for finite and for infinite $N$.

If the configurational microcanonical entropy $\omega_N^c(v)$ has a nonanalyticity at $v=v_c$, then the microcanonical entropy $\omega_N(\varepsilon)$ has a nonanalyticity at the same value $\varepsilon=v_c$ of its argument for any finite value of the number of degrees of freedom $N$. The presence of the kinetic energy weakens the nonanalyticities. More precisely, if the configurational entropy is $p$ times differentiable, then the entropy is $p+\lfloor N/2 \rfloor$-times differentiable. In the thermodynamic limit, however, the behaviour is very different: The nonanalyticities do not longer occur at the same values of the arguments, but the nonanalyticity of the microcanonical entropy is shifted to a larger energy as given by equation \eref{defestar}.

These results give a general explanation of what had previously been observed for the mean-field spherical model \cite{CaKa:06}. With the hypercubic model we have provided a further example for which both, the configurational microcanonical entropy and the microcanonical entropy can be computed analytically. Due to its simplicity, this model is of pedagogical value and nicely illustrates the general behaviour of nonanalyticities of the entropy.

The study of nonanalyticities of the configurational microcanonical entropy, their relation to stationary points of the potential energy, and their connection with topology changes of the constant-potential surfaces in configuration space, has attracted some interest recently and has proved useful to obtain a deeper understanding of phase transitions (see \cite{CaPeCo:00,Kastner:08,Pettini} for reviews). With the present work, we add another piece to this understanding by providing the relevant ``translation rules'' between configurational quantities and their total-energy counterparts. 

\section*{Acknowledgments}
L.C.\ thanks Physikalisches Institut, Universit\"at Bayreuth for hospitality during the early stage of this work.

\appendix
\section{Finite-$N$ nonanalyticities of the density of states: explicit calculation}
\label{appendix}

We now explicitly calculate the convolution integral (\ref{conv}) at finite $N$. For the sake of simplicity, let the configurational density of states $\omega_N^c(v)$ have just one nonanalyticity at $v = 0$ (this choice of the stationary value will ease the notation). According to the general results reported in \cite{KaSchneSchrei:08} and recalled in section \ref{sing}, we assume, apart from irrelevant multiplicative constants,
\beq
\omega^c_N(v) = 
\Theta\left(v - v_{\mathrm{min}}\right) \left[\tilde \omega^c_N(v) + \omega_\pm(v)\right] ,
\eeq
where the $\Theta$ distribution forces $\omega^c_N$ to vanish when the potential energy per degree of freedom $v$ is smaller than its minimum\footnote{Since we assume $v_c = 0$, we have $v_{\mathrm{min}} <0 $.} $v_{\mathrm{min}}$, $\tilde\omega^c_N(v)$ is smooth and $\omega_\pm(v) = {\cal O}\left( v^p\right)$ is the nonanalytic part; more precisely, to the leading order,
\beq
\omega_\pm(v) = \cases{
c^- |v|^p & \mbox{for $v < 0$},\\
c^+ |v|^p & \mbox{for $v > 0$},
}
\eeq
where $c^-,c^+\in\RR$ and $2p\in\NN_0$. With the the explicit expression \eref{eq:omegakin} for the kinetic density of states the convolution integral (\ref{conv}) becomes
\beq
\omega_N(\varepsilon) = \tilde\omega_N(\varepsilon) + a_N 
\int_0^{\infty} \rmd\gamma \, \gamma^{N/2-1} \omega_{\pm} 
(\varepsilon - \gamma) \,
\Theta\left(\varepsilon - \gamma - v_{\mathrm{min}}\right)
\eeq
with some smooth function $\tilde\omega_N(\varepsilon)$. We are interested in the integral containing the nonanalyticity,
\beq
I = \int_0^{\infty} \rmd\gamma \, \gamma^{N/2-1} \omega_{\pm} 
(\varepsilon - \gamma) \,\Theta\left(\varepsilon - \gamma - v_{\mathrm{min}}\right),
\eeq
and we can write $I=I_1+I_2$ with
\bea
I_1 & = & 
\int_0^{\infty} \rmd\gamma \, \gamma^{N/2-1} c^+ |\varepsilon - \gamma|^p\,
\Theta(\varepsilon - \gamma),\\ 
I_2 & = & 
\int_0^{\infty} \rmd\gamma \, \gamma^{N/2-1} c^- |\varepsilon - \gamma|^p\, 
\Theta(\gamma - \varepsilon)\, \Theta\left(\varepsilon - \gamma - v_{\mathrm{min}}\right). 
\eea
With the substitution $x = \gamma/\varepsilon$ we obtain
\beq
\fl I_1 = c^+\varepsilon^{p +N/2} \Theta(\varepsilon) \int_0^1 \rmd x\,x^{N/2-1} (1 -x)^p
 = c^+\varepsilon^{p +N/2} B(N/2,1+p) \, \Theta(\varepsilon),
\eeq
where $B(u,w) = \Gamma(u)\Gamma(w)/\Gamma(u+w)$ is the beta function. 

For the calculation of $I_2$ it is convenient to treat the cases $\varepsilon < 0$ and $\varepsilon > 0$ separately. For $\varepsilon > 0$, 
\beq
I_2 = c^-\int_\varepsilon^{\varepsilon - v_{\mathrm{min}}} \rmd\gamma\,
\gamma^{N/2-1} \left(\gamma - \varepsilon\right)^p,
\eeq
and with the change of variables $x = \varepsilon-\gamma$ one gets
\beq
I_2 = -c^-\int_0^{v_{\mathrm{min}}} \rmd x\,
(-x)^p \left(\varepsilon-x\right)^{N/2-1}.
\eeq
The substitution $y = x/\varepsilon$ yields
\bea
I_2 & = -c^-(-1)^p\varepsilon^{p +N/2} \int_0^{v_{\mathrm{min}}/\varepsilon} \rmd y\,y^p (1 -y)^{N/2-1} \nonumber \\
& = -c^- (-1)^p\varepsilon^{p +N/2} B_{v_{\mathrm{min}}/\varepsilon}(1+p,N/2),
\eea
where $B_z$ denotes the incomplete beta function.\footnote{The incomplete beta function $B_z$ has a branch cut discontinuity in the complex plane running along the negative real axis. This implies that, when expressing $I_2$ in terms of $B_z$, the correct branch has to be used. The given integral representations, however, are unambiguous.} For $\varepsilon < 0$, a similar calculation gives
\beq
I_2 = c^- (-1)^{-N/2}(-\varepsilon)^{p +N/2} B_{1-v_{\mathrm{min}}/\varepsilon}(N/2,1+p)\Theta(\varepsilon-v_{\mathrm{min}}).
\eeq
Assembling the pieces together we finally obtain for the nonanalytic part of the density of states
\beq
\label{omegasing}
\omega_N(\varepsilon) - \tilde\omega_N(\varepsilon) \propto 
|\varepsilon|^{p +N/2} \left[b^-(\varepsilon)\, \Theta(-\varepsilon) +
b^+(\varepsilon)\, \Theta(\varepsilon) \right],
\eeq
where
\bea
b^- (\varepsilon) & = & c^- (-1)^{-N/2} B_{1-v_{\mathrm{min}}/\varepsilon}(N/2,1+p)\Theta(\varepsilon-v_{\mathrm{min}}),\\
b^+ (\varepsilon) & = & c^+ B\left(N/2,1+p\right) -c^- (-1)^p B_{v_{\mathrm{min}}/\varepsilon}(1+p,N/2).
\eea
The function given by equation (\ref{omegasing}), and thus the entropy $s_N (\varepsilon)$, is nonanalytic at $\varepsilon = 0 = v_c$. At this point, $s_N$ is $\lfloor p + (N-1)/2 \rfloor$-times differentiable, whereas the configurational entropy $s_N^c$ we started out with is only $\lfloor p -1/2 \rfloor$-times differentiable. Without any modifications the calculation can be carried over to the integrated density of states $\Omega_N$.

\section*{References}

\bibliographystyle{unsrt}
\bibliography{kinsing.bib}

\end{document}